\documentclass{spie2004}
\usepackage[square,numbers]{natbib}
\usepackage{amssymb}
\usepackage{amsmath}
\usepackage[dvips]{epsfig}
\usepackage[dvips]{graphicx}

\def\##1{{\underline #1}}
\def\~#1{{\underline {\mathcal#1}}}
\def\+#1{{{\mathcal #1}}}
\def\=#1{\underline{\underline #1}}

\def\.{\mbox{ \tiny{$^\bullet$} }}

\begin{document}

\title{Porosity Effect on Surface Plasmon Resonance from Metallic Sculptured Thin Films}

\author{Ibrahim Abdulhalim,\supit{a}   Akhlesh Lakhtakia,\supit{b} Amit Lahav,\supit{a} Fan Zhang,\supit{b}
and Jian Xu\supit{b}
\skiplinehalf
\supit{a}Department of Electrooptic Engineering,
Ben Gurion University of the Negev, Beer Sheva 84105, Israel \\
\supit{b}Department of Engineering Science and
Mechanics, Pennsylvania State University, University Park, PA  16802, USA
}

\authorinfo{Further author information: (Send correspondence to I.A. E-mail: abdulhlm@bgumail.bgu.ac.il, Telephone:  972 8 646 1448}

\maketitle

\begin{abstract}  When a sculptured thin film (STF), made of a metal and $\leq50$~nm thick, is used in lieu of
a dense layer of metal in the Kretschmann configuration, experimental data for an STF comprising parallel
tilted nanowires shows that a surface plasmon
resonance (SPR) can still be excited. As the porosity of the chosen STF increases, experimental
data and numerical
simulations indicate the SPR dip
with respect to the angle of incidence of the exciting plane wave widens and eventually disappears,
leaving behind a a vestigial peak near the onset to the total-internal-reflection regime.

\end{abstract}


 \section{Introduction}
\begin{figure}[!ht]
\centering \psfull 
\epsfig{file=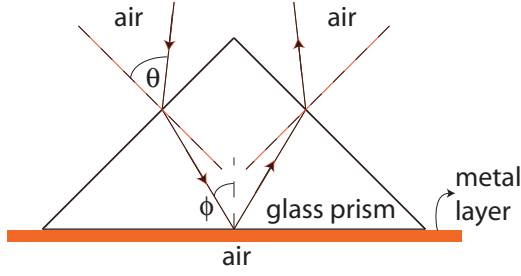,width=2.7in}
\caption{{\small 
Schematic of the Kretschmann configuration. $p$-polarized light  is incident on the
left slanted face of the $45^\circ$-$90^\circ$-$45^\circ$ glass prism at an angle
$\theta\in\left(-\pi/2,\pi/2\right)$. After partial refraction into the prism and partial reflection at the glass-metal interface,
it is partially refracted into air at the right slanted face of the prism. The fraction $R$ of the incident
power density that exits the right slanted face is plotted as a function of the angle $\theta$. A
sharp dip in $R$ as a function of $\theta$ indicates the resonant excitation of a surface plasmon wave at the
metal-air interface. The width and depth of the resonance is changed when a material is placed below the metal
 layer. The angle $\phi=(\pi/4)-\sin^{-1}\left( n_g^{-1}\,\sin\theta\right)$, where $n_g$ is the refractive
 index of glass.
 }} 
\label{Kretsch} 

\end{figure}

 The phenomenon of surface plasmon resonance  (SPR) at a planar metal-dielectric interface
 has been known for a long time \cite{Raether} and is widely exploited for
 optical sensing of chemicals \cite{HYG,AZLe}. In the  
 Kretschmann configuration, a thin layer of a metal
 is deposited on one face of a prism made of glass of high refractive
 index. As shown in Fig.~\ref{Kretsch},
 $p$--polarized light shining into the second face of the prism is partially refracted
 towards the metal layer and then partially reflected out of the third face of the prism.
 As the angle of incidence $\theta$ of light changes, the reflected light  exhibits a sharp dip 
 when the propagation constant along the metal-glass surface closely matches the 
 propagation constant of the surface plasmon wave guided by the metal-air interface. 
 This resonance---which does not occur with $s$--polarized
 light---occurs due to the real part of the relative permittivity of the metal
 being negative at optical frequencies. The decreased reflectance indicates high absorption
 in the metal layer. The width and depth of the resonance is changed when a material is placed below the metal
 layer. The material can either be the analyte to be sensed or a substrate containing that analyte.
 The accuracy of the SPR sensor can be enhanced by optimizing the thickness of the metal layer \cite{LCR}. 
 
Surface plasmons can also be localized---on nanoparticles and nanoshells \cite{KTGK} as well as
 at some locations in certain nanoporous structures \cite{MGbook}.  
The absorption resonances
associated with localized SPs   often show up in optical spectrums as narrowband features. Statistical fluctuations
of nanoscale morphology can lead to averaging over many different localized resonances, the overall
effect being broadened.  With this motivation, we decided to investigate the effect on localized SPR (LSPR) of
replacing the metal layer in the Kretschmann configuration by a sculptured thin film (STF) of a metal.

The nanostructure of an STF comprises clusters of 3-5 nm diameter. STFs can be made of inorganic or organic dielectrics, metals, and semiconductors, and deposited on variety of substrates, using physical vapor deposition techniques  such as thermal evaporation, electron-beam evaporation, and sputtering \cite{LMbook,Mattox}. STFs are assemblies of
naominally parallel and identical nanowires, all of the same shaped. When the nanowires are straight,
the STF is called a columnar thin film (CTF).
The nanowires are not single grains; thus, CTFs possess inter-columnar porosity as well as intra-columnar porosity,
and their nanoscale morphology is appropriate for the simultaneous excitation of multiple LSPRs. Indeed,
LSPR has been excited in porous metal thin films due to scattering
and therefore without the need for prism coupling, 
but only in the infrared regime \cite{MGSC}. We demonstrate here, for the first time,
 the excitation of LSPR using prism coupling on the surface of a porous aluminum thin film  in the visible regime.
 
\section{Experiments}
\begin{figure}[!ht]
\centering \psfull 
\epsfig{file=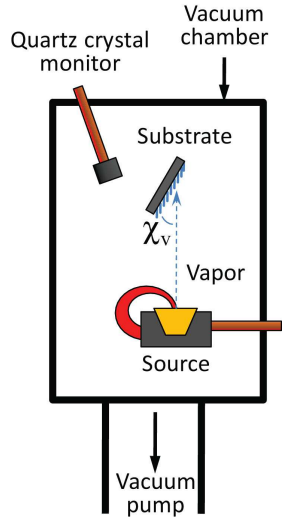,height=2.7in}
\caption{{\small 
Schematic of the oblique-angle-deposition technique. Collimated vapor flux oriented
at angle $\chi_v$ with respect to the substrate plane leads to the formation
of parallel columns tilted at an angle $\chi\geq\chi_v$ with respect to the same plane.}} 
\label{OAD} 
\end{figure}

CTFs of aluminum were deposited in an electron-beam evaporation system (PVD-75, KJL Inc.). The 
oblique-angle-deposition (OAD) technique was used, as depicted schematically in Fig.~\ref{OAD}. In an
evacuated chamber, with the vacuum base pressure set below $4$~$\mu$Torr, collimated aluminum vapor
was directed towards a 2.54-cm $\times$ 2.54-cm BK7-glass substrate at an angle $\chi_v=20^\circ$ to the substrate plane. The
distance between the aluminum source and the centroid of the substrate was set at $25.4$~cm. The deposition
rate was automatically controlled at $0.25$~nm~s$^{-1}$,  which was monitored with a resonating quartz crystal sensor. 
A profilometer (Tencor P-10) was used to measure the metal layer's average thickness as $30$~nm.

The Kretschmann configuration of Fig.~\ref{Kretsch} was implemented with incident light coming from
a 653-nm-wavelength laser diode collimated with a lens and then $p$-polarized after passage through
a linear polarizer. The prism was made of BK glass ($n_g=1.51509$). The glass substrate on which
the aluminum CTF was deposited was separated from the prism
 by a thin layer of an index--matching fluid. The fraction $R$ of the incident
power density that exits the right slanted face was measured using an amplifying photodiode. The entire
assembly was mounted on an optical table on which the angle $\theta$ can be set to $1^\circ$ precision.

\section{Numerical simulations}
The $R$-vs.-$\theta$ relationship was also numerically simulated in the following manner: The
$p$--polarization
transmittance $T_{ag}$ at the air-glass interface on the left side of the prism was computed as
a function of $\theta$ with the Fresnel formula \cite{BW}, as also the
$p$--polarization
transmittance $T_{ga}$ at the glass-air interface on the right side of the prism. The 
$p$-polarization reflectance $R_{gca}$ of glass-CTF-air at the base of the prism was computed
as a function of the angle $\phi=(\pi/4)-\sin^{-1}\left( n_g^{-1}\,\sin\theta\right)$, by
using a simple matrix-based formalism  wherein we assumed that the columns of the
CTF are tilted in the plane of Fig.~\ref{Kretsch} at an angle $\chi=70^\circ>\chi_v$ with respect to
the substrate plane \cite{LMbook}. The
effective relative
permittivity tensor of the aluminum CTF was obtained from a Bruggeman formalism \cite{SLH},
wherein it was assumed that aluminum is distributed in the form of electrically small
prolate spheroids of aspect ratios $1:1.2:15$, the void regions are electrically small
spheres, the void regions are vacuous, the porosity $p\in\left[0,1\right]$, and the relative permittivity
of bulk aluminum is $-59.288+22.2385i$. The CTF thickness was taken as $30$~nm. 
The quantity $R$
was \emph{estimated} as the product $T_{ag}R_{gca}T_{ga}$,
while the porosity $p$ of the CTF and the incidence angle $\theta$ were varied to fit
the experimental data.

\section{Results and discussion}
\begin{figure}[!ht]
\centering \psfull 
\epsfig{file=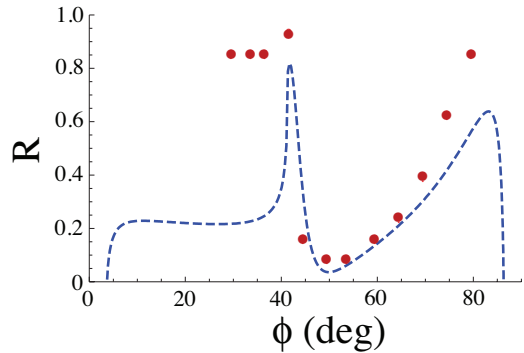,width=2.7in}
\caption{{\small 
$R$ as a function of $\phi$ when the metal layer is the
pre-etching aluminum CTF. The red dots indicate measured values,
while the dashed blue line indicates simulations made with $p=0.517$.
}} 
\label{517} 
\end{figure}

Figure~\ref{517} shows measured values of $R$ versus $\phi\in\left[35^\circ,80^\circ\right]$ for
an aluminum CTF. The sudden drop in $R$ as $\phi$ increases from $40^\circ$ to $45^\circ$
indicates the excitation of an SPR. The wide dip in the $R$--$\phi$ curve is indicative of scattering loss
due to the spatial nonuniformity of the distribution
of matter in the CTF. Fitting the numerically simulated  $R$-vs.-$\theta$ relationship
to the experimental data, we found that $p=0.517$; i.e., the CTF is 51.7\% volumetrically porous.

\begin{figure}[!ht]
\centering \psfull 
\epsfig{file=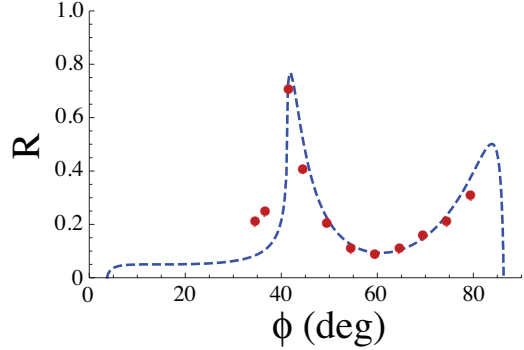,width=2.7in}
\caption{{\small 
$R$ as a function of $\phi$ when the metal layer is the
post-etching aluminum CTF. The red dots indicate measured values,
while the dashed blue line indicates simulations made with $p=0.560$.
}} 
\label{560} 
\end{figure}

Next, we etched the aluminum CTF  using the Transene AL etchant of type A for
$5$~s. The etching rate at 
$25$~$^\circ$C was 1~nm~s$^{-1}$. The CTF was then rinsed in water and dried by
blowing dry nitrogen on it.  Thereafter, $R$ was measured again for $\phi\in\left[30^\circ,80^\circ\right]$.
The measured data is presented in Fig.~\ref{560}. Clearly, the SPR dip is very different from that
in the previous figure. Numerical simulation of the $R$-vs.-$\theta$ relationship suggests that $p=0.56$.
In other words, etching increased the porosity from 51.7\% by 4.3\%, and the effect was captured by a
widening of the SPR dip.

\begin{figure}[!ht]
\centering \psfull 
\epsfig{file=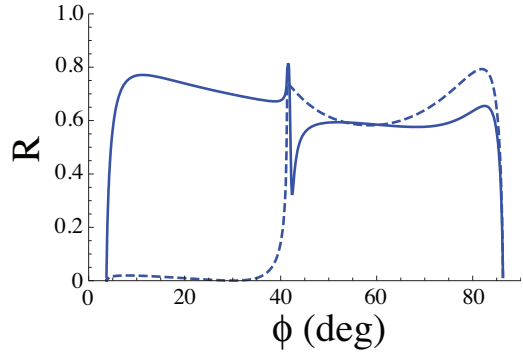,width=2.7in}
\caption{{\small 
Numerical simulations of
$R$ as a function of $\phi$ when $p=0.30$ (solid line) and $p=0.75$ (dashed line).
}} 
\label{fig5} 
\end{figure}

When sub-50-nm thick, CTFs prepared by obliquely directed physical vapor deposition are highly discontinuous, as they are made of nano-islands but with some orientational order which builds up more and more as columns as the thickness increases \cite{MY}.  As SPR in the Kretschmann configuration had been excited only with dense thin films, the
possibility of its excitation using metallic CTFs less than 50 nm thick was questionable.  We have now shown both theoretically and experimentally that SPR excitation is possible on sub-50-nm thick aluminum CTFs with porosity $\sim 50$\%.  The SPR dip widens  and becomes asymmetric as the porosity increases.   
As exemplified by Fig.~\ref{fig5},  numerical simulations indicate that, as the porosity increases beyond 75\%, the SPR dip almost disappears, with a vestigial peak near the onset to the total-internal-reflection regime.   

In conclusion, we expect the presented work to eventually lead to peak sensors with high sensitivity because 
a CTF is a porous material, and SPR excitation would be sensitive to variations
in its effective permittivity tensor due to infiltration of its void regions by a fluid \cite{Linf}. We 
also hope that this work points to a method to determine the porosity of CTFs, and of more general STFs also.

\end{document}